------------------------------------------------------------------------------

\documentstyle[12pt]{article}
\title{\bf  Gravitational Wave  Detector for a Space Laboratory}
\author{L.V.Verozub}
\begin{document}
\maketitle
\centerline{\em Department of Physics and Astronomy , Kharkov State
University, Kharkov 310077, Ukraine}
\begin{abstract}
     We propose a new  method of gravitational waves detection in the
$10^{-1}\div 10^{-2}$  - Hz band for a space laboratory based on the use of
the
Kozorez effect in the magnetic interaction of superconducting solenoids.
\end{abstract}
\section{Introduction}
     As Thorne noted \cite{Thorne}   it can be expected that  the  amplitude
$h$ of
gravitational wave bursts from astrophysical sources reaches the values
of the order of $10^{-16} \div 10^{-17}$ in $10{-1} \div 10^{-2}$ -Hz band .
Some authors \cite{Braginsky} proposed
to use  in this frequency  band the Doppler tracking of interplanetary
spacecraft , Skyhook-detector and the exitations of seismic motions in the
Earth`s surface  . A new type of the gravitational wave detector for the
low frequency band is cosidered in this paper. The detector is based on
the effect that the potential energy of a pair of magnetically interacting
superconducting solenoids in weightless state , generally speaking, has
the minimum at some nonzero distance. At this distance the solenoids are
in a weak  equilibrium  condition.  This  system in vacuum  is  a  weakly
coupled nonlinear oscillator  with very low energy dissipation and it can
be used as a sensitive detector of tidal low-frequency accelerations of
the solenoids. For measuring the tidal accelerations  of the order of
$10^{-13} cm/s^2$  in a satellite the gravity gradiometer \cite{Paik}
  was proposed. However,
the tidal acceleration of test bodies caused by low-frequency bursts is
of the order of  $10^{-16} \div 10^{-18}\, cm/s$ .
\section{ A Peculiarity  of Magnetically Interacting Superconducting
Solenoids}.

     Consider  a pair of the superconducting solenoids $P_{1}$  and $P_{2}$
in line, with points $x_{1}$  and $x_{2}$  as the centers in the weightless
state . If the solenoids carry the persistent currents $I_{1}$  and $I_{2}$  ,
their magnetic energies are  $U_{1} = I_{1} Q_{1} /2$  and
$U_{2} = I_{2} Q_{2} /2$,  where $Q_{1}$  and $Q_{2}$   are
magnetic fluxes in the solenoids. Since $Q_{1}  = L_{1} I_{1}  + M I_{2}$
and $Q_{2}  = L_{2} I_{2} + M I_{1}$   , where $L_{1}$   and $L_{2}$  are the
inductances and $M$ is the mutual inductance of the solenoids, the currents
are
\begin{equation} \label{I}
     I_{1}  = (Q_{1} L_{2} - Q_{2} M)/D ,\;
 I_{2}  = (Q_{2} L_{1}  - Q_{1} M)/D,
\end{equation}
where  $D = L_{1}L_{2} - M^2$  .The fluxes $Q_{1}$  and $Q_{2}$  are constants
and $M$ is a function of the distance $x = | x_{1}  - x_{2}|$ between the
solenoids
centers. Therefore, $I_{1}$  and $I_{2}$  are the function of $x$ and, when
one of the conditions
\begin{equation} \label{M}
 M = L_{1} Q_{2} /Q_{1}    , \; M = L_{2} Q_{1} /Q_{2}
\end{equation}
is satisfied, the currents change their signs.
    It follows  from  eqs.(\ref{I}) that the energy
$U = U_{1} + U_{2}$  of the system is given by
\begin{equation}  \label{U}
 U = ( L_{2} (Q_{1})^{2} - 2M Q_{1}Q_{2} - L_{1}(Q_{2})^{2} )/(2D).
\end{equation}

    The energy of the solenoids magnetic interaction is
\begin{displaymath}
  W = U - L_{1}Q_{1}^2 /2  - L_{2}Q_{2}^2 /2 .
\end{displaymath}
   The Ampere force $F = - \partial W / \partial x$ affects the solenoids.
It follows from the eq.(\ref{U})  that $F = I_1 I_2 \partial M /\partial x$ ,
where $I_1$  and $I_2$  are defined by eqs. (\ref{I}) .

    The basic peculiarity of the interaction of  superconducting
circulating currents (The Kozorez effect \cite{Kozorez} ) is that the
function
$W(x)$ has the minimum and the Ampere force is reduced to zero at the certain
distance $x = x_{0}$  , where one of the conditions (\ref{I}) is satisfied .

     Indeed, $F = U'_{M} M'_{x}$  ,where $ U'_{M} = \partial U / \partial M$
  and $ M'_{x}  = \partial M \partial x$ . However, the
function $U '_{M}  =[ Q_{1} Q_{2} M^2  + (L_{2} Q_{1}^2 + L_{1} Q_{2}^2) M -
Q_{1} Q_{2} L_{1} L_{2}] /D^2$    is
equal to zero if one of the conditions (\ref{M}) is satisfied.

     We have also $U''_{xx} = U''_{MM} (M'_{x})^2 + U'_{M}  M''_{xx}$   ,
where
$U''_{xx} = \partial U'_{x}/\partial x$ and $M'_{xx} =
\partial M'_{x}/\partial x$ .The mutual inductance $M(x)$ is a monotonically
 decreasing function with an increasing distance $x$ , and
$M'(x) \neq 0 $. Let $Q_{2} < Q_{1}$. Since $M < L_{1}$ , the first equation
 in (\ref{M}) is satisfied and
under this condition $U''_{MM}(x_{0} ) =
(Q_{1})^2 /[(L_{1} L_{2} - M^2) L_{1}] > 0$ .Then   $U''_{xx}(x_{0}) > 0$ .
 If $Q_{2} > Q_{1}$ , the second equation in (\ref{M}) is satisfied and we
again arrive at this conclusion. Therefore,the function $W(x)$ has a minimum
at the distance $x_{0}$ . If $Q_{2} < Q_{1}$ , the energy  $W$ in the
equilibrium position $x = x_{0}$  is $W_{min} = - Q_{2}^2 /(2L_{2})$.
     An illustrative example for the pair of the identical solenoids is
given in Fig. 1 . The parameters of the solenoids are : the inductances
are  $1.15 Hn.$ ,  the lengths are $5 cm.$, the radiuses  are  $10  cm.$,
$Q_{1} = 1.15 Wb$ ,$ Q_{2}  = Q_{1} / 25$ .

\setlength{\unitlength}{0.240900pt}
\ifx\plotpoint\undefined\newsavebox{\plotpoint}\fi
\sbox{\plotpoint}{\rule[-0.200pt]{0.400pt}{0.400pt}}%
\special{em:linewidth 0.4pt}%
\begin{picture}(1500,900)(0,0)
\tenrm
\put(264,473){\special{em:moveto}}
\put(1436,473){\special{em:lineto}}
\put(264,158){\special{em:moveto}}
\put(284,158){\special{em:lineto}}
\put(1436,158){\special{em:moveto}}
\put(1416,158){\special{em:lineto}}
\put(242,158){\makebox(0,0)[r]{-0.01}}
\put(264,315){\special{em:moveto}}
\put(284,315){\special{em:lineto}}
\put(1436,315){\special{em:moveto}}
\put(1416,315){\special{em:lineto}}
\put(242,315){\makebox(0,0)[r]{-0.005}}
\put(264,473){\special{em:moveto}}
\put(284,473){\special{em:lineto}}
\put(1436,473){\special{em:moveto}}
\put(1416,473){\special{em:lineto}}
\put(242,473){\makebox(0,0)[r]{0}}
\put(264,630){\special{em:moveto}}
\put(284,630){\special{em:lineto}}
\put(1436,630){\special{em:moveto}}
\put(1416,630){\special{em:lineto}}
\put(242,630){\makebox(0,0)[r]{0.005}}
\put(264,787){\special{em:moveto}}
\put(284,787){\special{em:lineto}}
\put(1436,787){\special{em:moveto}}
\put(1416,787){\special{em:lineto}}
\put(242,787){\makebox(0,0)[r]{0.01}}
\put(264,158){\special{em:moveto}}
\put(264,178){\special{em:lineto}}
\put(264,787){\special{em:moveto}}
\put(264,767){\special{em:lineto}}
\put(264,113){\makebox(0,0){0.15}}
\put(498,158){\special{em:moveto}}
\put(498,178){\special{em:lineto}}
\put(498,787){\special{em:moveto}}
\put(498,767){\special{em:lineto}}
\put(498,113){\makebox(0,0){0.2}}
\put(733,158){\special{em:moveto}}
\put(733,178){\special{em:lineto}}
\put(733,787){\special{em:moveto}}
\put(733,767){\special{em:lineto}}
\put(733,113){\makebox(0,0){0.25}}
\put(967,158){\special{em:moveto}}
\put(967,178){\special{em:lineto}}
\put(967,787){\special{em:moveto}}
\put(967,767){\special{em:lineto}}
\put(967,113){\makebox(0,0){0.3}}
\put(1202,158){\special{em:moveto}}
\put(1202,178){\special{em:lineto}}
\put(1202,787){\special{em:moveto}}
\put(1202,767){\special{em:lineto}}
\put(1202,113){\makebox(0,0){0.35}}
\put(1436,158){\special{em:moveto}}
\put(1436,178){\special{em:lineto}}
\put(1436,787){\special{em:moveto}}
\put(1436,767){\special{em:lineto}}
\put(1436,113){\makebox(0,0){0.4}}
\put(264,158){\special{em:moveto}}
\put(1436,158){\special{em:lineto}}
\put(1436,787){\special{em:lineto}}
\put(264,787){\special{em:lineto}}
\put(264,158){\special{em:lineto}}
\put(850,68){\makebox(0,0){$ x [m]$}}
\put(733,630){\makebox(0,0)[l]{$F(x) \; [N]$}}
\put(428,787){\special{em:moveto}}
\put(440,736){\special{em:lineto}}
\put(452,689){\special{em:lineto}}
\put(463,648){\special{em:lineto}}
\put(475,610){\special{em:lineto}}
\put(487,577){\special{em:lineto}}
\put(498,547){\special{em:lineto}}
\put(510,521){\special{em:lineto}}
\put(522,497){\special{em:lineto}}
\put(534,476){\special{em:lineto}}
\put(545,457){\special{em:lineto}}
\put(557,440){\special{em:lineto}}
\put(569,425){\special{em:lineto}}
\put(580,412){\special{em:lineto}}
\put(592,400){\special{em:lineto}}
\put(604,390){\special{em:lineto}}
\put(616,381){\special{em:lineto}}
\put(627,373){\special{em:lineto}}
\put(639,366){\special{em:lineto}}
\put(651,361){\special{em:lineto}}
\put(662,355){\special{em:lineto}}
\put(674,351){\special{em:lineto}}
\put(686,347){\special{em:lineto}}
\put(698,344){\special{em:lineto}}
\put(709,342){\special{em:lineto}}
\put(721,340){\special{em:lineto}}
\put(733,338){\special{em:lineto}}
\put(745,337){\special{em:lineto}}
\put(756,336){\special{em:lineto}}
\put(768,336){\special{em:lineto}}
\put(780,336){\special{em:lineto}}
\put(791,336){\special{em:lineto}}
\put(803,336){\special{em:lineto}}
\put(815,336){\special{em:lineto}}
\put(827,337){\special{em:lineto}}
\put(838,338){\special{em:lineto}}
\put(850,339){\special{em:lineto}}
\put(862,340){\special{em:lineto}}
\put(873,341){\special{em:lineto}}
\put(885,343){\special{em:lineto}}
\put(897,344){\special{em:lineto}}
\put(909,345){\special{em:lineto}}
\put(920,347){\special{em:lineto}}
\put(932,348){\special{em:lineto}}
\put(944,350){\special{em:lineto}}
\put(955,352){\special{em:lineto}}
\put(967,353){\special{em:lineto}}
\put(979,355){\special{em:lineto}}
\put(991,357){\special{em:lineto}}
\put(1002,359){\special{em:lineto}}
\put(1014,360){\special{em:lineto}}
\put(1026,362){\special{em:lineto}}
\put(1038,364){\special{em:lineto}}
\put(1049,366){\special{em:lineto}}
\put(1061,367){\special{em:lineto}}
\put(1073,369){\special{em:lineto}}
\put(1084,371){\special{em:lineto}}
\put(1096,373){\special{em:lineto}}
\put(1108,374){\special{em:lineto}}
\put(1120,376){\special{em:lineto}}
\put(1131,378){\special{em:lineto}}
\put(1143,379){\special{em:lineto}}
\put(1155,381){\special{em:lineto}}
\put(1166,383){\special{em:lineto}}
\put(1178,384){\special{em:lineto}}
\put(1190,386){\special{em:lineto}}
\put(1202,387){\special{em:lineto}}
\put(1213,389){\special{em:lineto}}
\put(1225,390){\special{em:lineto}}
\put(1237,392){\special{em:lineto}}
\put(1248,393){\special{em:lineto}}
\put(1260,395){\special{em:lineto}}
\put(1272,396){\special{em:lineto}}
\put(1284,398){\special{em:lineto}}
\put(1295,399){\special{em:lineto}}
\put(1307,400){\special{em:lineto}}
\put(1319,402){\special{em:lineto}}
\put(1331,403){\special{em:lineto}}
\put(1342,404){\special{em:lineto}}
\put(1354,405){\special{em:lineto}}
\put(1366,407){\special{em:lineto}}
\put(1377,408){\special{em:lineto}}
\put(1389,409){\special{em:lineto}}
\put(1401,410){\special{em:lineto}}
\put(1413,411){\special{em:lineto}}
\put(1424,412){\special{em:lineto}}
\put(1436,414){\special{em:lineto}}
\end{picture}

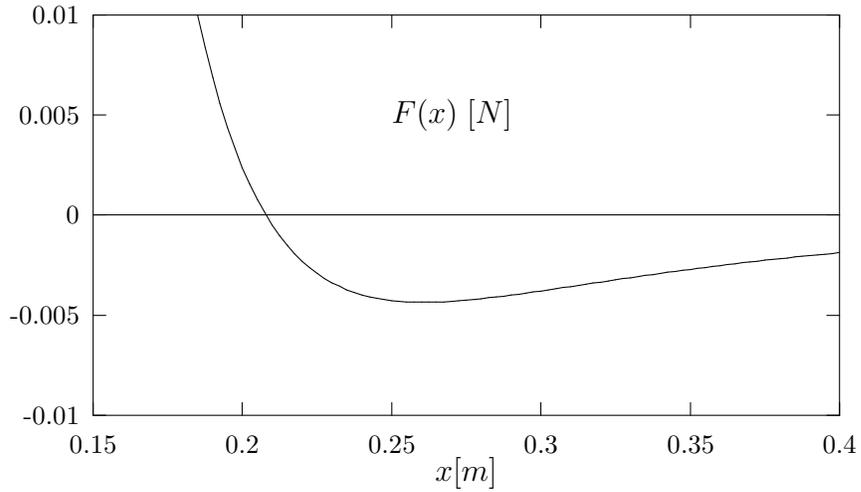
Fig.1 . The Force F(x)
\\
\\
\section{ The Magnetically Coupled Solenoids as a Detector of Tidal
 Accelerations}

     Consider the properties of the system, formed by a pair of
superconducting solenoids in weightless state, in the field of a
gravitational wave with the frequency $\nu$  extending during  the time
interval $t_{0}$
perpendicularly to $P_{1}P_{2}$ - direction. This system can be regarded as an
oscillator, the small oscillations of which in $P_{1} P_{2}$  - direction are
described by the nonlinear differential equation
\begin{equation} \label{eqoscillator}
     m\ddot{q} + R(\dot{q}) + pq = f(t) .                            
\end{equation}
     In eq.(\ref{eqoscillator})  $q = x - x_{0}$  is a small deviation from
the equilibrium position , $\dot{q} = \partial q/ \partial t$ ,
$\ddot{q}  = \partial \dot{q} / \partial t$ , $R(\dot{q})$ is the air
resistance, $ p = U'' (x_{0} )$
is the stiffness of the "magnetic spring",
$f(t) = m  a_{g} \sin (\omega t)$ at $0 < t < t_{0}$   and $f(t) = 0$  at
$t > t_{0}$  , $m$ is  the mass of the system , $\omega  = 2\pi \nu $
and $a_{g}  = \omega ^2 h x_{0}/2 $   is the amplitude of the tidal
acceleration, caused by the gravitational wave.

 If 1-type superconductors are used in the solenoids, the air resistance
$R(\dot{q})$ is the key cause of the oscillations damping in the given
system. (This assertion is argued in Section 4). Such a detector can be
named an ideal detector.For an ideal gas the function
$R(\dot{q})=-b\dot{q} |\dot{q}|$, where
 $b = \alpha \rho S/2$ , $\alpha $ is the aerodynamic factor of the
resistance , $\rho $ is the air density in the device, $S$ is the
crosssection of the solenoids orthogonally to the $P_{1} P_{2}$  - direction.
If the pressure is $10^{-10} Torr$ in the hydrogen atmosphere, the
temperature  $T = 4.2 K$ and $S = 10 cm$ , the magnitude of $b$ reaches
$10^{-14} gm/cm$ .

     Since $R(\dot{q})$ and $a_{g}$  are small quantities , we shall use the
 Bogolubov- Krilov method \cite{Blaquier} for the analysis of  nonlinear
equation (\ref{eqoscillator}).

Suppose, $\omega$  is close to the resonant frequency
$\omega_{0} = (p/m)^{1/2}$. We shall
seek an approximate solution of  eq.(\ref{eqoscillator}) in the form
\begin{equation}  \label{q(t)}
          q(t) = A(t)\cos [\omega t + \vartheta] ,          
\end{equation}
where $A$ and $\vartheta$ are slowly varied functions of time $t$. Let us
replace $R(\dot{q})$ in eq. (\ref{eqoscillator}) by the linear function
$\lambda _{e} \dot{q}$ , where $\lambda$  minimises the function
\begin{equation} \label{intmin}
  I(\lambda )= \int_{0}^{T}  [ R(\dot{q}) - \lambda_{e} \dot{q}] dt 
\end{equation}
$(T = 2 \pi) $. It is attained at $\lambda _{e} = \alpha b A \omega$, where
 $\alpha  = 8/(3\pi)$. Denote
$\lambda _{e}/m$   by $\beta $ . Then eq. (\ref{eqoscillator}) takes the form
of a linear differential equation
\begin{equation}  \label{eqlinoscil}
  \ddot{q} + \beta \dot{q} + (\omega _{0} )^2 q = a_{g} sin(\omega t) ,
\end{equation}
where, however, $\beta $ is a function of $A$.

    It follows from eq.(3) that at $Q_{1}  > Q_{2}$  the stiffness $p$ is
given by
\begin{equation} \label{stiffness}
       p =  [Q_{1} M'_{x} (x_{0})]^2 / (L_{1} D)            
\end{equation}
The typical values of $\omega$   are much less than $1 Hz$.
     Substituting eq. (\ref{q(t)}) into (\ref{eqoscillator}), ignoring the
terms $\ddot{A}$ and $\beta \dot{A}$ and
setting the coefficients of $\cos(\omega t)$ and $\sin(\omega t)$ equal to
zero, we obtain
the differential equations system for $A$ and $\vartheta$ :
\begin{equation} \label{eqForA}
   \dot{A} + \frac{\alpha \beta \omega ^{2} A}{2m} +
\frac{a_{g}}{2 \omega }\cos( \vartheta) = 0          
\end{equation}
\begin{equation} \label{eqForVartheta}
\dot{\vartheta} + \frac{ \omega ^{2} -\omega _{0}^{2}}{2 \omega  m} -
  \frac{a_{g}}{2Am} \sin(\vartheta) = 0    
\end{equation}

   Setting $\dot{A} = 0$ and $\dot{\vartheta} = 0$  and eliminating the phase
 $\vartheta$ , we obtain the equation
\begin{equation}  \label{equation }
 A \bigl[ ( \omega ^2 -\omega _{0}^2 )^2 +
\alpha b^2  \omega ^4 m^{-2} A^2  \bigr] = (\omega ^4 h x_{0})/4      
\end{equation}

     This equation gives the implicit function  $A = A(\omega )$ at stationary
oscillations. At the resonance ($ \omega = \omega_{0}$ ) the amplitude is given
 by
\begin{equation}  \label{amplitudeA}
            A =  \bigl [ mhx_{0} / (2\alpha b) \bigr ]^{1/2} .  
\end{equation}
At $\omega  = \omega _{0}$  and zero initial condition there  exists solution
$A = A(t)$ with a constant phase $\vartheta = \pi$ :
\begin{equation}  \label{A}
            A(t) = A_{m} \tanh (t / \tau _{0} ) ,             
\end{equation}
where  $ \tau _{0} = \omega ^{-1} [ 8m/(\alpha bhx_{0}]^{1/2}$
is the detector relaxation time .If in eq. (\ref{amplitudeA}) $m = 10^4 gm$ ,
$x  = 50 cm$ , $b = 10^{-3}  gm/cm$ , $h = 10^{-22}$   and
$\omega  = 0.1 Hz$  we find that the stationary amplitude of the detector
response $A_{m}  = 1.5*10^{-7} cm$.  However , it is impossible to observe
such high amplitudes since relaxation time of the ideal detector
( $\tau _{0}$  is proportional to $h^{-1/2}$    )
is too high. However,  $A(t)$ reaches  $ 2 \cdot 10^{-15} cm.$
already at the  observation time $t = 2.6 \cdot 10^6  s$ (one month).

   At $t \ll t_{0}$  we obtain $A(t)=(A_{m} / \tau _{0} ) t$. Consequently,
if the interval of a resonant gravitational-wave burst is $T = 2\pi / \omega$,
 the detector response is
\begin{equation}   \label{q(t)att>t0}
          q(t) = (\pi h  x_{0} /2 ) \cos (\omega _{0} t)   
\end{equation}
at $t > T$. Thus, the gravitational wave resonant burst gives rise to the
long - duration , poorly damped oscillations with the frequency $\omega$
and the amplitude  $ \pi h x_{0} / 2 $ . At $h = 10^{-17} $    and
$x_{0}  = 50 cm$  the amplitude is $7.8\cdot 10^{-16} cm$.

    If the resonant frequency $\omega _{0}$  is much less than $\omega$  ,
then a numerical solution of eq.(\ref{q(t)}) shows that the response
$q(t)$ reaches  $1.7*10^{-15} cm$ at $t = 10 s$  and further on  slightly
varies in time .

\section{ Measurement noises}
    The detector under study is a nonlinear oscillator. Let us find the
variance of its thermal fluctuations that is other than the one of a linear
oscillator.
    Consider first the linear oscillator describred by the differential
equation of the form  $\ddot{q} + \beta \dot{q} + \omega _{0}^2 q = f(t)$.
Suppose,  the  oscillator relaxation time is much longer than the
measurements time, i.e. $\beta t\ll 1$.
In that case , acording to a rigorous solution by Chandrasekhar
\cite{Chandrasekhar}  the
position $q$ variance $(\sigma _{q})^2$    is given by
\begin{equation} \label{sigmaq}
   \sigma _{q}^2   = \frac{kT}{m\omega ^2}\beta t
\biggl[1 - \frac{\sin(2\omega _{0}t)}{2\omega _{0} t} \biggr ] + o(\beta t^2)
\end{equation}
and the velocity $v = \dot{q}$  variance is given by

\begin{equation}  \label{sigmav}                            
    \sigma _{v}^2 = \frac{kT}{m} \beta t
\biggl [ 1 + \frac{\sin(2\omega _{0} t)}{2\omega t} + o(\beta  t^2) \biggr ]
\end{equation}

     The expressions between  the brackets are essential at low resonant
frequencies of the oscillator.

   In the case under consideration for an approximate analysis of the
thermal fluctuations it is expedient to take the statistic linearization
of the function  $R(\dot{q})$. Namely, we put
$R(\dot{q}) = \lambda _{1e} \dot{q}$ , where $\lambda _{1e}$   minimises
the function  $I_{1} (\lambda _{1e}) =
\langle [ R(\dot{q}) - \lambda _{1e} \dot{q}] \rangle $ .
( The symbol $\langle \; \rangle$ denotes
an ensemble mean). Now the random oscillations of the detector, caused by
the thermal noise, are given approximately by the differential equation
$\ddot{q} + b \dot{q} + \omega ^2 q = \xi(t)$, where
$\beta  = \lambda _{1e}/m$
and $\xi(t)$ is a Gaussian noise.

 The desired  value of $\lambda _{1e}$ is  $\lambda_{1e}  =
< \dot{q}  R(\dot{q}) > / < \dot{q}^2 >$ .

    Assuming  that the distribution function is approximately Gaussian,
we find  that $\beta  = 4(2\pi)^{-1/2} m^{-1} \sigma _{v}$ . Thus, $\beta$
is the function of $\sigma _{v}$  .

    Now we are considering eq. (\ref{sigmav}) as an equality whence the
function $\sigma _{v} (t)$ can be found. Substituting the above expression of
$\beta$ into eq.(\ref{sigmav})
and found function $\sigma _{v}(t)$ - into  eq.(\ref{sigmaq}) we obtain,
finally, that the  value  of   the variance  $\sigma _{q}^2$ : .
\begin{equation}
\sigma_{q}^2 (t) = \frac{4b^2}{ (2 \pi)^{1/2} m \omega _{0}^2}
\frac{kT}{m} t^2
\biggl [ 1 - \frac{\sin^2 (2 \omega _{0} t )}{2 \omega _{0}^2 t)^2} \biggr ]
\end{equation}                                                      

     For example, if $m = 10^4  gm$ , $\nu _{0}  = 0.01 s$, $t = 100 s$ ,
$b = 10^{-4} gm/cm$,
then the root-mean-square magnitude of the thermal fluctuations  is :
$< q^2 >^{1/2} = \sigma _{q} = 10^{-20} cm$. Thus, the detector has a  very
low level of the
thermal noise, that is much less than the expected detector response to
the gravitationalwave bursts.

     The inhomogeneity of  the  Earth and spacecraft gravitational fields
lead to  more serious problems. The difference in the solenoids gravitati-
onal accelerations approximately equal to $U_{ik}\; x_{0}$ , where
$U_{ik} = \partial^2{U} /\partial x^i \partial x^k $
 and $U$ is the Earth gravitational potential. The variations of $U_{ik}$
   during  the orbital motion of the spacecraft are well beyond the tidal
accelerations  $a_{g}$   in  eq.(4). This problem can be
solved
by choosing a geostationary or very distant from the Earth spacecraft orbit.

 It is necessary to take into account the fluctuations in the gravity
gradient within the spacecraft too. For example, the $ 1 cm $ shift in the
position of the $5 gm$ mass at its distance $2 m$ from the solenoids along
the $P_{1} P_{2}$  -direction causes the variation $4 \cdot 10^{-14} cm/s^2$
in the relative acceleration of the solenoids.

 A little part of the inductance (the motional inductance) is a function
of the temperature $T$. So, variations in $T$ cause the variations
$\delta L$ in the inductance $L$. (In the above illustrative example at the
wire radius $0.05 cm$  $dL/L$ is about $10^{-16}$ at $T = 2 K$ and $10^{-20}$
at $T = 0.1K$ ).

    For definiteness, assume that $Q_{1} < Q_{2}$  . Then under  the
equilibrium position $Q_{2}L_{1} - Q_{1} M = 0$  and $D \approx  L_{1} L_{2}$
. If at  moment $t=0$ a temperature variation begins, then, according to
eqs. (2), at $t > 0$ we have
$I_{1} \approx I_{1}^0 + I_{1}^0 (\delta L_{2} /L_{2}) $ and
$I_{2} \approx I_{2}^0 (\delta L_{1}/L_{1})$   ,where
$I_{1}^{0} = Q_{1}/ L _{1}$  and $I_{2}^0 =Q_{2} /L_{2}$.
Now the force of the interaction between the solenoids in absence of exterior
 forces is other than zero :
\begin{equation}  \label{force}
  F_{T}  = I_{1}^{0} I_{2}^{0} (\delta L_{1} /L_{1}) M'(x_{0} ) ,   
\end{equation}
where$ M'(x_{0}) = \partial M / \partial x $ . The force $F_{T}$ gives rise
to slow oscillations of the solenoids relative to the equilibrium position
$x_{0}$ .

    Denote the acceleration $F_{T} /m$ by $a_{T}$  . Obviously, it is
necessary to select the detector parameters so that during the measurements
time the inequality  $a_{g}  < a_{T}$  is valid. We do not analyze the
problem of optimization of the detector parameters  here. However, it
should be noted that it is easiest of all to obey the above inequality if the
detector resonant frequency $\omega_{0}$  is much less than the frequency
$\omega$   of the wave . This happens to be the case for the considered above
illustrative example  (Fig.1) at $Q_{1}  = 1.15 \cdot 10^{-1}  Wb$.  In that
case  $\omega_{0}  = 2.6\cdot 10^{-3} s^{-1}$  , $I_{1}^{0}  = 0.1 A$
and $I_{2}^0 = I_{1}^0 /25$ ). At $T = 2K$  and $\delta L/L = 10^{-16}$
the acceleration $a_{T} = 5\cdot  10^{-18} cm/s$ . Meanwhile, at
$\nu  = 0.1 Hz$ and $h = 10^{-17}$  the gravitational acceleration
$a_{g} = 4 \cdot 10^{-17} cm/s $. (The detector response to  short bursts
found by the the numerical solution of eq.(4) is about $7\cdot 10^{-16} cm$
at $t_{0}  =10 s$).

    It follows from eq.(8) that the stiffness
$p = [ I_{1}^{0} M '(x_{0} )]^2 /L_{2}$.
Hence, the temperature fluctuations cause  the resonant frequency
variations $\delta \omega _{0}$ . They are given by
$\delta \omega _{0} /\omega _{0} = - \delta L_{1} /(2 L_{1} )$.These
variations
may be essential at the resonant detection of a continuous signal.

   Under oscillations  of  the  solenoids  relative  to  the  equilibrium
position  the persistent currents in the solenoids are not constants
( $ \dot{I} = dI/dt \neq 0 $ ).  From the viewpoint of the  two-fluid  model
the intensity of the supercurrent is $J = en_{s}v_{s}$  , where $e$ is the
charge of the electron, and $n_{s}$  and $v_{s}$  are the concentration and
velocity of the superconducting electrons ,correspondingly. As a result of
the acceleration of the superconducting electrons  an electric field $E$
appears inside the superconducor that can be found from the equality
$m\dot{v}_{s} = - eE$ , where  $\dot{v}_{s}  = dv_{s}  /dt$. If the
time -depending current is of the form  $J \exp(-i\omega t)$ , then $ E  =
-[i m\omega J/(e^{2} n_{s} )]$. This electric field causes a motion of the
normal electrons according to the following equation \cite{VanDuzer}
\begin{displaymath}
m d \langle v_{n} \rangle /dt + (m/\tau_{0} ) \langle v_{n} \rangle = -e E,
\end{displaymath}
where
$\langle v_{n} \rangle$ is the mean
velocity of the normal electrons and $\tau_{0}$  is
the relaxations time (usually about $ 10^{-13} s$). Then, there is also a
normal current in the solenoids  with the intensity
$J_{n} = - e n_{n} \langle v_{n}\rangle $ , where $n_{n}$  is the normal
electrons concentration. Solving the above equation of motion  we find that at
low frequencies $\omega$ the normal current is given by

\begin{equation} \label{normCurrent}
          I_{n} = (n_{n} /n_{s} )\omega \tau I_{s}^0 \sin(\omega t) .  
\end{equation}

    Let $Q_{1}  < Q_{2}$  . The amplitude $I_{2}^{max}$ of the supercurrent
in the solenoid $P_{2}$  caused by gravitational-wave bursts is equal to
$ I_{2}' A$ , where $A$ is the amplitude of the detector response and
$I_{2}' = dI_{2}/dq$ . It follows from eq.(1) that
$I_{1}' = (I_{2}^{0}/L_{1}) M'$    and $I_{2} ' = (I_{1}^{0} /L_{2}) M'$  .The
 typical value of  $I_{2}'$  is of the order of $10^{-2} \div 10^{-4} A/cm$.
 Since $A$ is about $10^{-16} cm$,  the typical value of $I_{s}^{max} $
 proves to be about $10^{-19} A$ . Setting in (19)  $n_{n}/ n_{s} = 1/ 5$
 and $\nu  = 0.1 Hz$ we find that the amplitude of the normal current is
$10^{-34} A$. During time $dt$ the energy dissipation is $RI_{n}^2 dt$,
where $R$ is the superconductor wire normal resistance . If $R = 1 \Omega $ ,
 then during the time $T = 2\pi /\omega $ the energy dissipation is about
$10^{-67} J$ . At the same time the tidal forces work is about
$m a_{g} A = 10^{-36} J$ .

 Consider  another question:  ought we to include in eq.(10) in addition
to the term  $R(\dot{q})$ an effective force  $F_{ef}$ , which describes
the energy dissipation caused by the normal current ? In  principle, it can
be done since  according to eq.(19) $I_{n}$  is of the form
$(n_{n} /n_{s})\tau \dot{I}_{s} (t)$. Hence,
$RI^2 dt$ can be written as $ F_{ef} \dot{q} dt$ , where
$F_{ef}= R \tau ^2 (\partial I / \partial q)^2 \dot{q}$ . Let
us compare $F_{ef}$  with the force $R(\dot{q})$. The value of
$\dot{q}$ is approximately $ \omega A $ .
At low frequencies and at $b \gg 10^{-10} gm/cm$  the following inequality
is fulfilled: $R(\dot{q}) \ll F_{ef}$   , since
$R \tau ^2 (I')^2 \ll bA\omega$ . (Because of too longe relaxation time
$\tau _{0}$  , small values of $b$ ought not to be  used ).

    Thus, at very small amplitude oscillations and low frequencies the
energy dissipation caused by the normal current is insignificant.

     Consider briefly the physical principles of the solenoids relative
shift measurement caused by the gravitational-wave bursts.
     For definiteness, assume that $Q_{2} < Q_{1}$  . At the equilibrium
position the supercurrent $I_{2} = 0 $. The deviations in the equilibrium
position cause the change in the proper magnetic flux of the solenoid $P_{2}$
is $\delta Q = \delta I_{2} L_{2}$,
where $ \delta I_{2}  = I'_{2} q $ . Thus, the value $\delta Q_{2}$ is given
by

\begin{equation}   \label{dQ2}
  \delta Q_{2} = I_{1}^0 M ' q             
\end{equation}

    Suppose that the distance between the near solenoids ends is less than
$1 cm$ at the distance between the solenoids centers of $50 cm$ . In that
case the value $M '$ reaches $10^{-2} Gn/cm$ or more than that. At
$I_{1}^0 = 0.1 A$  $ dQ_{2}$
is about $10^{-3} \Phi _{0}$  , where $\Phi _{0} = 2\dot 10^{-15} Wb$ is
the magnetic flux quantum.

    A method of $dQ_{2}$  measuring is shown in Fig.2 . In this schematic
diagram $S$ is a superconducting quantum- interferometer device (SQUID) that
is attached to the solenoid $P_{2}$  and coupled with $P_{2}$  inductively
by a flux transformer $T$.

\setlength{\unitlength}{0.240900pt}
\ifx\plotpoint\undefined\newsavebox{\plotpoint}\fi
\begin{picture}(1500,900)(0,0)
\tenrm
\put(264,158){\special{em:moveto}}
\put(1436,158){\special{em:lineto}}
\put(264,158){\special{em:moveto}}
\put(264,787){\special{em:lineto}}
\put(264,158){\special{em:moveto}}
\put(1436,158){\special{em:lineto}}
\put(1436,787){\special{em:lineto}}
\put(264,787){\special{em:lineto}}
\put(264,158){\special{em:lineto}}
\put(411,473){\makebox(0,0)[l]{$P_{1}$ }}
\put(704,473){\makebox(0,0)[l]{$P_{2}$}}
\put(967,309){\makebox(0,0)[l]{$T$}}
\put(1260,284){\makebox(0,0)[l]{$S$}}
\put(293,208){\special{em:moveto}}
\put(293,284){\special{em:lineto}}
\put(323,410){\special{em:lineto}}
\put(352,284){\special{em:lineto}}
\put(381,410){\special{em:lineto}}
\put(411,284){\special{em:lineto}}
\put(440,410){\special{em:lineto}}
\put(469,284){\special{em:lineto}}
\put(498,410){\special{em:lineto}}
\put(528,284){\special{em:lineto}}
\put(557,410){\special{em:lineto}}
\put(586,284){\special{em:lineto}}
\put(586,208){\special{em:lineto}}
\put(293,208){\special{em:lineto}}
\put(645,208){\special{em:moveto}}
\put(645,284){\special{em:lineto}}
\put(674,410){\special{em:lineto}}
\put(704,284){\special{em:lineto}}
\put(733,410){\special{em:lineto}}
\put(762,284){\special{em:lineto}}
\put(791,410){\special{em:lineto}}
\put(821,284){\special{em:lineto}}
\put(850,410){\special{em:lineto}}
\put(879,284){\special{em:lineto}}
\put(909,410){\special{em:lineto}}
\put(938,284){\special{em:lineto}}
\put(967,271){\special{em:lineto}}
\put(938,259){\special{em:lineto}}
\put(967,246){\special{em:lineto}}
\put(938,233){\special{em:lineto}}
\put(967,221){\special{em:lineto}}
\put(938,208){\special{em:lineto}}
\put(645,208){\special{em:lineto}}
\put(1026,284){\special{em:moveto}}
\put(997,271){\special{em:lineto}}
\put(1026,259){\special{em:lineto}}
\put(997,246){\special{em:lineto}}
\put(1026,233){\special{em:lineto}}
\put(997,221){\special{em:lineto}}
\put(1026,208){\special{em:lineto}}
\put(1202,208){\special{em:lineto}}
\put(1231,221){\special{em:lineto}}
\put(1202,233){\special{em:lineto}}
\put(1231,246){\special{em:lineto}}
\put(1202,259){\special{em:lineto}}
\put(1231,271){\special{em:lineto}}
\put(1202,284){\special{em:lineto}}
\put(1026,284){\special{em:lineto}}
\put(1260,221){\special{em:moveto}}
\put(1260,259){\special{em:lineto}}
\put(1319,259){\special{em:lineto}}
\put(1319,221){\special{em:lineto}}
\put(1260,221){\special{em:lineto}}
\end{picture}

Fig.2
\\
\\
The minimal magnetic flux measured by the SQUID
with the $T$ is given by \cite{squid}
$ \delta Q_{min} = 2 \bigl [2 L_{a} {\cal E} (\nu ) \bigr ]^{-1/2} / N_{a}$,
where $L_{a}$  is the
antenna inductance , $N_{a}$  is its turns number and ${\cal E}(\nu )$ is the
SQUID noise expressed as an input energy resolving at the frequency $\nu $ .
At the
frequencies $\nu \leq 0.1 Hz$  the value ${\cal E}(\nu ) =
( 10^{-31} / \nu n ) J/(Hz)^{-1/2}$     .  At
$\nu = 0.1 Hz$  $\delta Q_{min} = 10^{-4}  \Phi _{0}$  or less than that.
Thus, $\delta Q_{min} < \delta Q_{2}$  .

      We mean that the solenoids $P_{1}$  and $P_{2}$  are inside a
supeconducting
shield.     The    results  of   the   SQUID   measurements  are
transmitted by radio by means of a conversion "voltage - frequency" and
by using  an isotropic active antenna. Such a method of the solenoids
shift measuring is insensitive to micrometeorites impacts and other forces
affecting the spacecraft.

     In the SQUID circuit there  exists noise magnetic flux $ \Phi _{n}$
of the order of $10^{-5} \Phi _{0}$  or less than that. Because of the
inductive coupling between the SQUID and the solenoid $P_{2}$   a noise
current appears in the
latter that is less than $\Phi / L_{2} = 10^{-20} A$ . It is much less than
the
noise effect upon the detector is a small value.

     A number of the other disturbing forces in the space laboratory is
considered in [11 ]  .
     The author would like to thank  C.W.F. Everitt and P.W.Worden Jr.
for the materials on the STEP.

\end{document}